\documentclass[twocolumn,aps,pre,showpacs,ams,amssymb]{revtex4}
\usepackage{graphicx}% Include figure files
\usepackage{dcolumn}% Align table columns on decimal point
\usepackage{bm}% bold math
\def\vep{\varepsilon}

\def\bfr{{\bf r}}
\def\nab{{\bf \nabla}}

\begin{document}
\title{A practical method to estimate
the condensate fraction of interacting and trapped Bose atoms}
%%%%%%%%%%%%%%%%%%%%%%%%%%%%%%%%%%%%
\author{Sang-Hoon \surname{Kim}\footnote{shkim@mmu.ac.kr}}
 %\footnotesize
\affiliation{Division of Liberal Arts,
Mopko National Maritime University, Mopko 530-729, R.O. Korea}
 %%%%%%%%%%%%%%%%%%%%%%%%%%%%%%%%
 \date{\today}
\begin{abstract}
We suggest a practical way to estimate the condensate fraction of an
interacting dilute Bose gas confined by an external harmonic
potential as a function of temperature and scattering length. It
shows that an increase of the scattering length produces an exponential
 decrease of condensate fraction.
\end{abstract}
\maketitle
%%%%%%%%%%%%%%%%%%%%%%%%%%%%%%%%%%%%%%%%%

Over the last decade
 the phenomenon of Bose-Einstein condensation of
ultra-cold dilute alkali Bose gas confined in a magneto-optic trap
 has been a subject of fundamental interest. Even below the
transition temperature some atoms leave the lowest energy state and
remain in the excited state by an atom-atom interaction. This causes
a depletion of the condensate number in comparison to its ground
state occupation for the non-interacting case, and it depends on
temperature and interaction strength both.

The study of the condensate fraction (CF), $N_0/N$, where $N$ being
total number of atoms, of a dilute Bose system has a long history.
The first microscopic theory was suggested by Bogoliubov. If the
system is homogeneous and dilute ($\rho a^3 \ll 1$) where $\rho$ is
the density, the CF at $T=0K$ is known from the Bogoliubov theory as
$ 1 - \frac{8}{3 \sqrt{\pi}} \sqrt{\rho a^3}$ \cite{huan,fett}.
However, this formula is not applicable for an inhomogeneous system
because confinement makes the wave function of the Bosons
space-varying and so is the density inhomogeneous.

In this paper we suggest a practical way to estimate the CF of
interacting Bosons in a harmonic trap. Our approach is to start with
a non-interacting harmonically trapped Bosons at $T \neq 0$ and to
extend the argument for the interacting case. Some experimental
variables will be applied for the practical purpose.

Consider a three-dimensional(3D) system of non-interacting bosons
confined by a harmonic potential of angular frequency ${\bf
\omega}$. Then, the number distribution function in the grand
canonical ensemble at temperature $T$ is given by
\begin{equation}
N= \sum_{n_x,n_y,n_z} \frac{1}{e^{\beta(\vep_n -\mu)}-1},
\label{10}
\end{equation}
where $\beta=1/k_B T$ and $\mu$ is the chemical potential. $\vep_n$
is the discrete energy spectrum of the harmonic oscillator given by
$\vep_{n}= \hbar(n_x\omega_x+n_y\omega_y+n_z\omega_z) + \vep_0, $
where $n_x,n_y,n_z=0,1,2,...$ and $\vep_0$ is the zero point energy.

We rewrite  Eq. (\ref{10}) as a sum of two parts: ground state and
excited states as
\begin{eqnarray}
N &=& \frac{1}{e^{\beta(\vep_0 -\mu)}-1} + \sum_{n=1}^{\infty}
\frac{d(n)}{e^{\beta(\vep_n -\mu)} -1} \nonumber
\\ &  \equiv & N_0^{non} + N_e,
\label{20}
\end{eqnarray}
where  $d(n)=(n+1)(n+2)/2$ is the degeneracy of the harmonic
oscillator. If we consider  an isotropic trap of
$\omega_x=\omega_y=\omega_z=\omega$, the number of atoms in the
condensate at temperature $T$ is well-known as
\cite{gros,haug1,haug2}
\begin{equation}
N_0^{non}(T)  = N - \left( \frac{k_B T}{\hbar\omega}\right)^3 g_3(z)
- \frac{3}{2}\left( \frac{k_B T}{\hbar\omega}\right)^2 g_2(z),
\label{30}
\end{equation}
where $z$ is the effective fugacity defined by
$z=e^{\beta(\mu-\vep_0)}$ and $g_s$ is the Bose series function
defined by $ g_s(z) = \sum_{n=1}^\infty z^{n}/n^s $. Eq. (\ref{30})
is solved numerically to obtain $N_0^{non}(T)$. Note that the
transition temperature for $N=10^4$ trapped atoms with $\omega=10^3
sec^{-1}$ is about $150 nK$.

 In the interacting system of
the interaction energy
 in the ground state $U(N_0,a)$,
 the zero point energy is shifted
 from $\vep_0$ to $\vep_0 + U$.
 Then, it depends on number of condensate atoms
 $N_0$ and the $s$-wave scattering length $a$, both.
 In this scheme,
the new ground state occupation
number
 can be rewritten as
\begin{equation}
N_0(T,a) = \frac{1}{e^{\beta(\vep_0 -\mu +U)}-1}.
\label{50}
\end{equation}

The ground state information of a dilute Bose gas is obtained from
the solution of Gross-Pitaevskii equation  as \cite{park,dalf}
\begin{equation}
i\hbar \frac{\partial  \Psi_0}{\partial t} = \left[
-\frac{\hbar^2}{2m} \nab^2 +\frac{1}{2}m\omega^2 r^2 +  U_0 |
\Psi_0(\bfr) |^2 \right] \Psi_0(\bfr,t), \label{52}
\end{equation}
where $U_0 (=4 \pi \hbar^2 a/m)$ is the interaction strength between
atoms and $\Psi_0$ is the macroscopic wave function of the
condensate which is normalized to the number of particles in the
ground state, $N_0$.
The macroscopic wave function  $\Psi_0$ is obtained from the
solution of Eq. (\ref{52}) numerically and  then, the ground state
interaction energy $U(N_0,a)$ is given by \cite{park,dalf}
\begin{equation}
U(N_0,a) = \int \frac{U_0}{2} | \Psi_0(\bfr) |^4  d^3r. \label{54}
\end{equation}
We plotted the interaction energy in Fig. 2 as a function of a
dimensionless interaction strength $g(N_0,a)=4\pi N_0 a/a_{ho}$,
where  $a_{ho}$ is the harmonic oscillator length given by
$a_{ho}=\sqrt{\hbar/m\omega}$. The $U$ is not linear
 but  close to a logarithmic function for small $g$
\cite{gnan}.

%%%%%%%%%%%%%%%%%%%%%%%%%%%%%%%%%%%%%%%%%%%%%%%%%%%
%%%%%% %%%%%%%%%      FIGURE 2    %%%%%%%%%%%%%%
%%%%%%%%%%%%%%%%%%%%%%%%%%%%%%%%%%%%%%%%%%%%%%%%%%%
\begin{figure}
\includegraphics{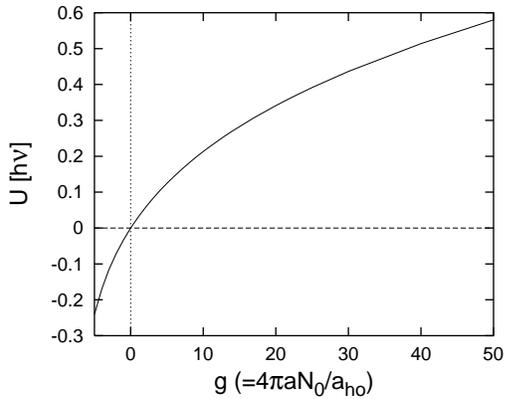}\\% Here is how to import EPS art
\caption{Interaction energy per atom of harmonically trapped and
interacting Bose atoms.
 The unit is $\hbar\omega$. }
\end{figure}
%%%%%%%%%%%%%%%%%%%%%%%%%%%%%%%%%%%%%%%%%%%%%%%%%

A direct calculation of Eq. (\ref{50}) is not reliable because Eq.
(\ref{30}) is too sensitive to get an accurate $z$. Instead,
comparing Eq. (\ref{50}) with Eq. (\ref{20}), we may obtain a
practical form of the CF as
\begin{equation}
N_0(T,a)  \sim  N_0^{non}(T) e^{-\beta U(N_0,a)}.
\label{55}
\end{equation}
This approximation is valid only when $|\beta U| \ll 1$.
 In the non-interacting limit
$N_0 \rightarrow N_0^{non}$,
 and otherwise  $N_0$
 falls exponentially as $U$ increases.

%%%%%%%%%%%%%%%%%%%%%%%%%%%%%%%%%%%%%%%%%%%%
%%% %%%%%%%%%      FIGURE 3    %%%%%%%%%%%%%%
%%%%%%%%%%%%%%%%%%%%%%%%%%%%%%%%%%%%%%%%%%%%
\begin{figure}
\includegraphics{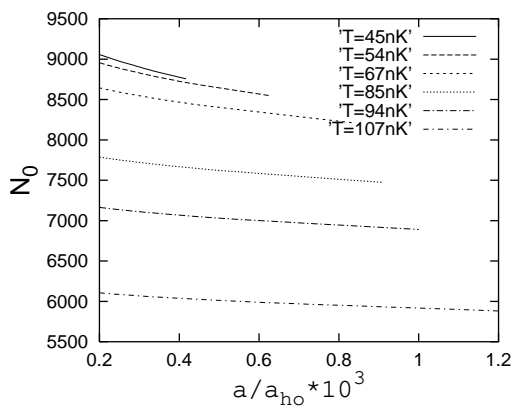}
\caption{Condensate number $N_0$
 as a function of scattering length % $a/a_{ho}$
 shown for various temperatures.}
\end{figure}
%%%%%%%%%%%%%%%%%%%%%%%%%%%%%%%%%%%%%%%%%%%%%%%

Because $U$ itself is a function of $N_0$, the CF has to be obtained
self-consistently. We obtain $N_0^{non}(T)$ from Eq. (\ref{30}) and
then put it into $g(N_0,a)$  to solve the Eq. (\ref{52})
numerically. Then, substituting the solution of Eq. (\ref{52}) into
Eq. (\ref{54}) again,
 we obtain the interaction energy $U$ as a function of
$N_0$ and $a$.
Next, substitution of $U$ into  Eq. (\ref{55}) again,
we obtain a new $N_0$.
 Then, it is put back into Eq. (\ref{52}) for self-consistency.

Calculations are carried out for $|\beta U| < 0.1$ with the
practical variables of $N=10^4$ and   $\omega=10^3 sec^{-1}$.
%Note that it corresponds to  $\beta U = 7.64 U[\hbar \omega]/T[nK]$.
 The results are plotted in Fig. 3.
The $N_0$ shows an exponential decrease as scattering length $a$
increases.
 The slope is steeper for lower temperatures.
 As scattering length increases
with other parameters remaining fixed, the depletion of the
condensate is enhanced.

We suggested a practical way to estimate the condensate fraction of
interacting and trapped Bose atoms in specific experimental ranges.
The effect of the
attractive interaction is not clear because it may increase CF a
little bit for negative scattering length compared with
non-interacting system. That is, there exists a possibility an
attractive interaction between fermions and bosons may increase CF
compared with that of bosons only.

%%%%%%%%%%%%%%%%%%%%%%%%%%%%%%%%%%%%%%%%%%
The author thanks Mukunda P. Das and G. Gnanapragasam for useful
discussions.
%%%%%%%%%%%%%%%%%%%%%%%%%%%%%%%%%%%%%%%%%%
{}
%%%%%%%%%%%%%%%%%%%%%%%%%%%%%%%%%%%%%%%

\begin{thebibliography}{}
\bibitem{huan} K. Huang, Statistical Mechanics, 2nd. (Wiley, New York, 1987) p. 335.
\bibitem{fett} A.L. Fetter and J.D. Walecka, Quantum Theory of Many-Particle Systems,
 (McGraw-Hill, New York, 1971) p. 221.
 \bibitem{gros} S. Grossmann and M. Holthaus, Phys. Lett. A {\bf 208}  188 (1995).
\bibitem{haug1} H. Haugerud, T. Haugset and F. Ravndal,
Phys. Lett. A {\bf 225} 18 (1997) .
\bibitem{haug2} T. Haugerud, H. Haugset and J. O. Anderson,
\pra {\bf 55}  2922 (1997).
\bibitem{park} A.S. Parkins and D.W. Walls, Phys. Rep. {\bf 303} 1  (1998).
\bibitem{dalf} F. Dalfovo, S. Giorgini, L.P. Pitaevskii, and S. Stringari,
        \rmp {\bf 71} 463  (1999).
\bibitem{gnan} G. Gnanapragasam, S.-H. Kim, and M.P. Das,
    Mod. Phys. Lett. B {\bf 20}  1839 (2006).
%%%%%%%%%%%%%%%%%%%%%%%%%%%%%%%%%%%%%%%
\end{thebibliography}
\end{document}